\documentclass[10pt,journal,a4paper]{IEEEtran}

\usepackage{cite}
\usepackage{url}
\usepackage{amssymb,amsmath}
\usepackage{array}
\usepackage{graphicx}
\usepackage{subfigure}

\begin{document}

\title{Relay Selection with Network Coding in Two-Way Relay Channels}

\author{\authorblockN{Yonghui~Li,~\IEEEmembership{Member,~IEEE,}
                      Raymond H. Y. Louie,~\IEEEmembership{Student~Member,~IEEE,}
                      and~Branka~Vucetic, ~\IEEEmembership{Fellow,~IEEE}}
\thanks{Copyright (c) 2010 IEEE. Personal use of this material is permitted. However, permission to use this material for any other purposes must be obtained from the IEEE by sending a request to pubs-permissions@ieee.org.}
\thanks{This work was supported in part by the Australia Research Council under Grant DP0985140, DP0877090 and LP0991663.}
\thanks{Manuscript received March 2 2010,  revised June 20 2010 and  accepted July 28 2010. The review of this paper was coordinated by Dr. Chau Yuen.}
\thanks{The authors are with the Centre of Excellence in Telecommunications, School of Electrical \& Information Engineering, University of Sydney, Sydney, NSW 2006, Australia. Email: \{lyh, rlouie, branka\}@ee.usyd.edu.au, Tel:+61 2 9351 2236, Fax: 61 2 9351 3847.}}

\maketitle

%

\begin{abstract}
In this paper, we consider the design of joint network coding (NC)
and relay selection (RS) in two-way relay channels. In the proposed
schemes, two users first sequentially broadcast their respective
information to all the relays. We propose two RS schemes, a single
relay selection with NC and a dual relay selection with NC. For both
schemes, the selected relay(s) perform NC on the received signals
sent from the two users and forward them to both users. The proposed
schemes are analyzed and the exact bit error rate (BER) expressions
are derived and verified through Monte Carlo simulations. It is
shown that the dual relay selection with NC outperforms other
considered relay selection schemes in two-way relay channels. The
results also reveal that the proposed NC relay selection schemes
provide a selection gain compared to a NC scheme with no relay
selection, and a network coding gain relative to a conventional
relay selection scheme with no NC.
\end{abstract}


\section{Introduction}
Wireless channels typically suffer from time varying fading caused
by multipath propagation and Doppler shifts, resulting in serious
performance degradation. Diversity has been an effective technique
in combating channel fading. Recently, a new form of diversity
technique, called user cooperative diversity \cite{1}, has been
proposed for wireless networks. The idea is to allow users to
communicate cooperatively by sharing their antennas to achieve a
spatial diversity gain. The use of relay aided transmission is one
example of a practical cooperative diversity technique. In a relay
system, the source sends its information to the relays. The relays
then process the received signals, and forward them to the
destination. At the destination, by properly combining the received
signals sent from the source and relays, cooperative diversity can
be achieved. It has been shown that cooperative communications can
dramatically improve the system capacity and performance \cite{4,5}.

To further improve the network capacity, the application of network
coding (NC) \cite{2} in wireless relay networks has recently drawn
significant attention. In particular, NC has been studied in
multiple access, multicast and two-way relay channels, where two
users communicate with each other with the help of relays
\cite{6,7,8,9,10}. Some physical layer NC schemes, joint
network-channel coding and scheduling algorithms, etc, have been
proposed \cite{6,7,8,9,10},\cite{16,17,18,19,20,21}. It has been
shown that properly designed NC can achieve significant capacity
improvement in cooperative wireless networks.

Most of the current work on two-way relay channels considers the use
of a single relay node to aid communication in the system
\cite{6,7,8,9,10}. In this paper, we consider a two-way relay system
with multiple relay nodes. In multiple relay networks, if all relays
participate in the relayed transmission, it is usually assumed that
they transmit on orthogonal channels so that they do not cause
interference to each other \cite{4,5}. Relaxing the orthogonality
constraint can lead to a capacity increase with an increased system
complexity. To overcome these problems, relay selection algorithms
using various relay protocols, such as amplify and forward, decode
and forward (DAF), and their variations, have been proposed to
facilitate system design for one way non-orthogonal multiple relay
networks \cite{11,12,13},\cite{28}. A commonly used relay selection
strategy in one way relay networks is to select a single best relay,
which has the optimal end to end performance or capacity among all
relays \cite{11,12,13}, or among all relays whose received
signal-to-noise ratios (SNRs) are larger than a threshold \cite{28}.
It was shown that the single relay selection can achieve the full
spatial diversity order as if all relays are used. Furthermore, the
system bit error rate (BER) performance and capacity compared to
all-participation relaying schemes is improved \cite{4,5}.

In this paper, we consider the design of relay selection for two-way
relay channels. In \cite{22}, an interesting relay selection scheme
was proposed for two-way relay channels. The relay selection
criterion was to maximize the weighted sum rate for any
bidirectional rate pair on the boundary of the achievable rate
region. It was shown that the probability that there exists one
relay node which achieves the optimal rate pair decreases with
increasing the number of relay nodes. The optimal relay selection
criterion decides for any rate pair individually and the optimal
rate region can be achieved by time-sharing of different relay
nodes.

In this paper, we propose practical relay selection schemes for
two-way relay channels, designed to minimize the average sum bit
error rate (BER) of the two end users in two-way relay channels. We
consider a decode and forward relaying protocol for information
forwarding. To improve the spectral efficiency and error performance
of bidirectional relayed transmission, we combine relay selection
(RS) and network coding, and develop efficient joint relay selection
and network coding schemes (RS-NC). Specifically, we propose two
RS-NC schemes, referred to as the single relay selection with NC
(S-RS-NC) and dual relay selection with NC (D-RS-NC) schemes. In the
proposed schemes, both users first sequentially send their
respective information to all the relays. Based on certain selection
criteria, a single relay or two relays are selected for
transmission. The selected relay(s) decode each user's signals,
perform NC and then broadcast the signals to both users. To
facilitate the analysis and selection process, we propose some
simple selection criteria for both relay selection schemes.
Specifically, for the single relay selection, inspired by the fact
that the overall error probability of two users is dominated by the
worst user, we propose a near-optimum \textit{Min-Max} selection
criterion for the S-RS-NC scheme, such that the BER of the worst
user among the two users is optimized. That is, a single relay,
which minimizes the instantaneous BER of the worst of the two users,
will be selected in the proposed S-RS-NC scheme. Simulation results
confirm that the \textit{Min-Max} selection criterion achieves
almost exactly the same performance as the optimal single relay
selection criterion based on the minimization of the average sum BER
of two users. During the revision of this paper, we have been
informed that a similar \textit{Min-Max} selection criterion for
two-way multiple antenna relay network has been proposed in
\cite{29,30}, but no closed-form BER expressions were derived. In
this paper, we will derive the exact closed-form and asymptotic BER
expressions, and the analytical results are verified by Monte-Carlo
simulations.

On the other hand, as revealed by the information-theoretic results
in \cite{22} that the bidirectional communication is characterized
by a two-dimensional rate pair, the optimal relaying may be achieved
by time-sharing of multiple relays. Motivated by this result, in
this paper we propose another dual-relay selection scheme, referred
to as the D-RS-NC. For the optimal D-RS-NC scheme, the destination
needs to do an exhaustive search to find the optimal pair of two
relays. This procedure is very complex, and requires coordination
between the two end users, and also involves significant amounts of
feedback. To facilitate the selection process, we propose a simple
selection criterion, named \textit{Double-Max} criterion. In this
criterion, we select one best relay for each user. Here, the best
relay for the user $k$, $k$=1, 2, means the relay, which has the
best link quality to the user $k$. The best relays for two users
could be the same or different. If the best relays for two users are
the same, then only a single relay is selected for transmission.
Otherwise, two different relays will be used. The proposed selection
scheme is very simple and can be easily implemented.

The performance of the proposed RS-NC schemes is analyzed and
verified by Monte Carlo simulations. Results show that both RS-NC
schemes can achieve the full diversity order as if all relays are
used. Results also show that the dual relay selection is superior to
the single relay selection and outperform other considered relay
selection schemes in two-way relay networks. This is different from
the conventional one way relay networks \cite{11,12,13} where the
single relay selection is the optimal selection strategy. We also
compare the performance of RS-NC schemes with other conventional
schemes. Results show that the combined relay selection and NC
provides a selection gain compared to the pure NC scheme with no
relay selection, and a network coding gain relative to the
conventional relay selection with no NC scheme. This implies that a
properly combined network coding and relay selection can improve
both system performance and spectral efficiency.

The rest of the paper is organized as follows. The system model is
described in Section II. In Section III, we propose several relay
selection schemes for two-way relay channels. The performances of
these schemes are analyzed and compared with the conventional
schemes. The results are verified by simulations in Section IV. In
Section V, we draw the conclusions.

\section{System Model}

In this paper, we consider a general two-hop two-way relay network,
where user 1 and user 2 exchange their information with the help of
$N$ relays. For simplicity, in this paper, we consider the BPSK
modulation. The extension to other modulation schemes is
straightforward.

Let $b_1(k)$ and $b_2(k)$ represent the $k$-th information bit
transmitted by user 1 and 2, respectively, and $s_1(k)$ and $s_2(k)$
denote the corresponding modulated symbols. The overall transmission
can be divided into three steps. In the first two time slots, user 1
and user 2 send their respective information symbol $s_1(k)$ and
$s_2(k)$ to all relays. The corresponding received signal at the
$i$-th relay transmitted from user $j$, denoted by
$y_{{u_j},{r_i}}(k)$, $j$=1, 2, $i$=1, 2,$\cdots$, $N$, can be
expressed as
\begin{equation} \label{eq1}
y_{{u_j},{r_i}}(k)=\sqrt{\bar{p}_{u_j}}h_{{u_j},{r_i}}(k)s_{j}(k)+n_{{u_j},{r_i}}(k)
\end{equation}
where $\bar{p}_{u_j}$, $j$=1, 2, is the average transmission power
at user $j$ and $h_{{u_j},{r_i}}(k)$ is the fading coefficient
between user $j$ and relay $i$. In this paper, we assume that all
fading coefficients are modeled as zero-mean, unit variance,
independent circular symmetric complex Gaussian random variables.
Furthermore, $n_{{u_j},{r_i}}(k)$ is a zero mean complex Gaussian
random variable with a noise variance of $\sigma_n^2$. In this
paper, we assume that all noise processes have the same variance
without loss of generality.

In the final step, the relays process the received signals and
transmit it to the two users. Specifically, the relays decode the
received signals from users 1 and 2, and perform network coding by
XORing the two decoded bits $b_1(k)$ and $b_2(k)$. Let $b_r(k)$
represent the binary summation of $b_1(k)$ and $b_2(k)$, and
$s_r(k)$ denote the corresponding modulated symbol. Let $x_{r,i}(k)$
be the $k$-th signal transmitted from the $i$-th relay, expressed as
\begin{equation} \label{eq2}
x_{r,i}(k)=\sqrt{\bar{p}_{r,i}}s_r(k)
\end{equation}
where $\bar{p}_{r,i}$ is the average transmission power at relay
$i$. Relay $i$ then broadcasts this signal to both users 1 and 2.
The corresponding received signal at the $j$-th user receiver can be
written as
\begin{equation} \label{eq3}
y_{{r_i},{u_j}}(k)=h_{{r_i},{u_j}}(k)x_{r,i}(k)+n_{{r_i},{u_j}}(k)
\end{equation}
where $h_{{r_i},{u_j}}(k)$ is the fading coefficient between relay
$i$ and user $j$.

After receiving signals from the relays, each user then decodes the
received signals and estimates $b_r(k)=b_1(k)\oplus {b_2(k)}$. Let
$\hat{b}_{r,1}(k)$ and $\hat{b}_{r,2}(k)$ represent the estimated
$b_r(k)$ at user $1's$ and user $2's$ receivers, respectively. Since
user $i$ already knows its own transmitted bit $b_i(k)$, it can
recover other user's bit by simply XORing $\hat{b}_{r,i}(k)$ with
its own bit $b_i(k)$. For example, user 1 can obtain an estimation
of $b_2(k)$, denoted by $\hat{b}_2(k)$, by performing
\begin{equation} \label{eq4}
\hat{b}_2(k)=b_{r,1}(k)\oplus b_1(k).
\end{equation}

\begin{figure}
\centering
\includegraphics[width=0.4\textwidth]{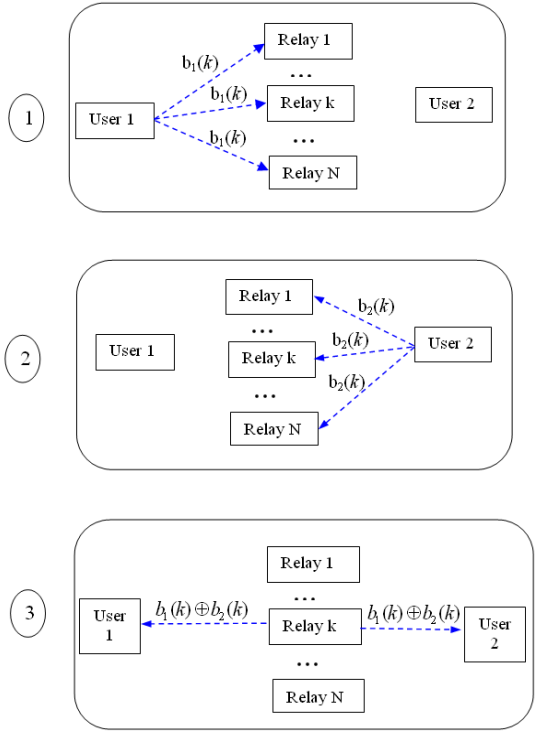}
\caption{Three step transmission of single relay selection with
network coding (S-RS-NC) in two-way relay networks} \label{fig:1}
\end{figure}

\section{Relay Selection With Network Coding In Two-Way Relay Channels}

In this section, we propose several joint relay selection and
network coding (RS-NC) schemes for two-way relay channels. We
consider a decode and forward relaying protocol, where the relays
decode the received signals before forwarding them to the
destination. Similar to \cite{25}-\cite{27}, in this paper, we focus
on the scenario where the link from the source to relay is much more
reliable than the link from the relay to destination and as such the
overall error probability from the source to the destination is
dominated by the error probability from the relay to the destination
and the contribution of decoding errors at the relay to the overall
performance can be ignored. For simplicity of analysis, we assume
error-free decoding at the relays. This is representative of many
practical scenarios, such as when the source transmission power is
large compared to the relay transmission power. One example includes
two base stations that exchange their information through a relay
station.

To facilitate the BER proofs in this section, note that after
integration by parts, we have
\begin{equation} \label{eq5}
E_x\left[Q\left(\sqrt{bX}\right)\right]=\frac{\sqrt{b}}{2\sqrt{2\pi}}\int_{0}^{\infty}\frac{e^{-bx/2}F_X(x)}{\sqrt{x}}dx
\end{equation}
where $Q(\bullet)$ is the $Q$ function and  $F_X(x)$ is the
cumulative distribution function (CDF) of $X$. In addition, if the
first order expansion of the probability density function (PDF) of
$X$ can be written in the form
\begin{equation} \label{eq6}
f_X(x)=\frac{ax^N}{\gamma^{N+1}}+o\left(x^{N+\epsilon}\right),\epsilon>0
\end{equation}
at high SNR, the asymptotic BER is given by \cite{24}
\begin{eqnarray} \label{eq7}
E_x\left[Q\left(\sqrt{bX}\right)\right]&=&\frac{2^Na\Gamma(N+3/2)}{\sqrt{\pi}(N+1)}(b\gamma)^{-(N+1)}
\nonumber \\ &+& o\left(\gamma^{-(N+1)}\right)
\end{eqnarray}

\begin{figure}
\centering
\includegraphics[width=0.49\textwidth]{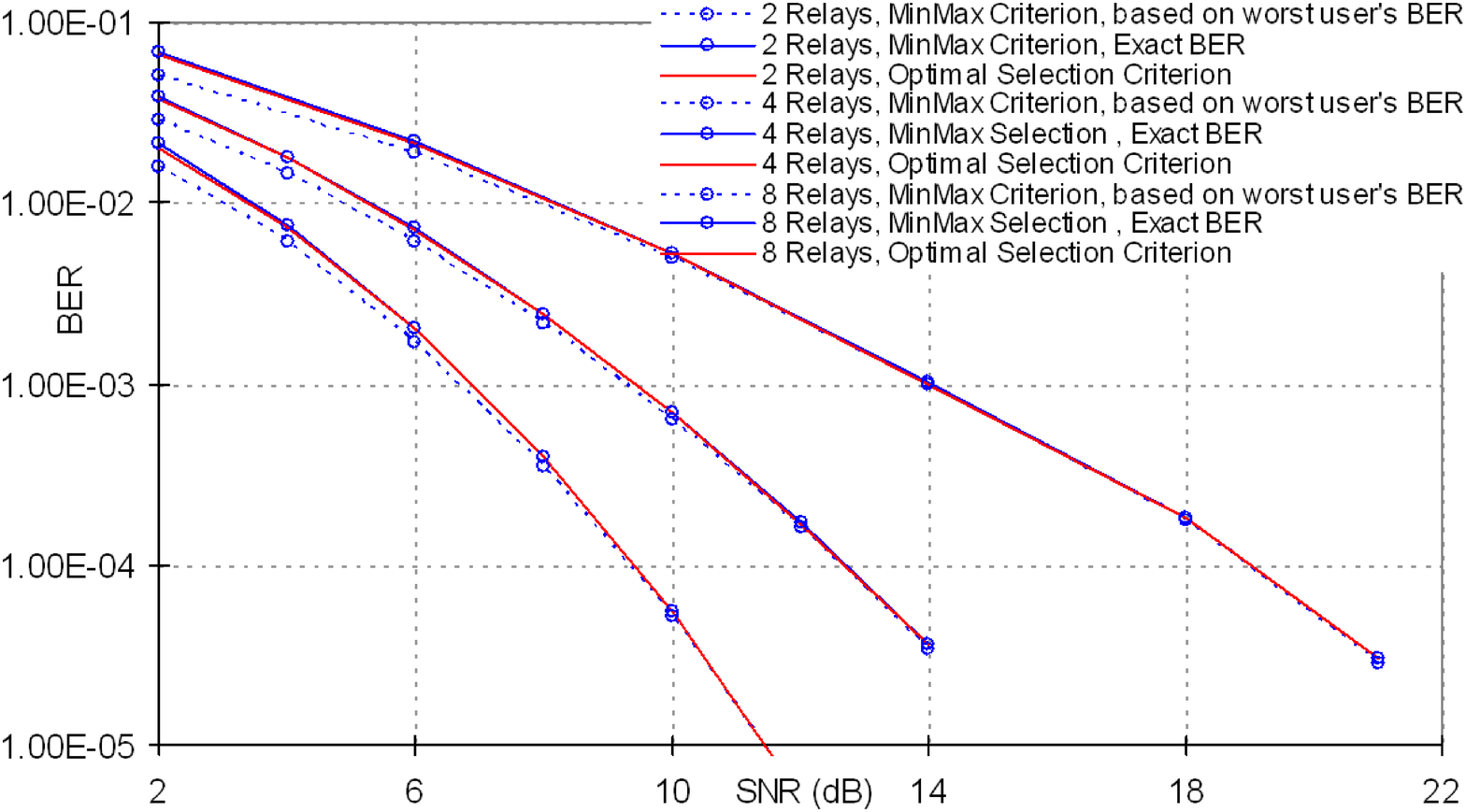}
\caption{The comparison of BER based on worst user's BER and exact
BER using MinMax selection criterion and the exact BER based on the
optimal selection criterion for single relay selection with NC}
\label{fig:2}
\end{figure}

\subsection{Single Relay Selection with Network Coding (S-RS-NC) }

In this section, let us first consider an RS-NC scheme, referred to
as the single RS-NC (S-RS-NC) scheme, where only a single relay,
which optimizes the system performance, is selected. Its block
diagram is shown in Fig. 1. For different selection criteria, the
selection decision procedures are different, so we will discuss it
separately for each scheme.

Realizing the fact that the average sum BER of two end users is
dominated by the worst user, in this paper, we consider a simplified
selection criterion for which the instantaneous BER of the worst
user among two users is minimized. We refer to such a selection
criterion as the \textit{Min-Max} selection criterion. Fig. 2
compares the performance of an approximate BER using only the worst
user's BER and exact two users' BER based on the \textit{Min-Max}
selection criterion, as well as the exact two users' BER based on
the optimum selection criterion for which the average sum BER of two
users is minimized. To facilitate the simulation of the optimal
selection criterion, we use a Chernoff bound to approximate the
$Q$-function when determining the optimal relay. In this figure, we
assume that the average SNR in the link from relay $i$ to user $j$
for $i=1,\ldots,N$ and $j=1, 2$, are all equal. As can be seen from
the figure, the worst user's BER is almost the same as the exact two
users' BER at medium to high SNR regions. We can also observe that
the \textit{Min-Max} selection criterion achieves almost exactly the
same BER performance as the optimal relay selection at all SNR
range. This confirms that the \textit{Min-Max} selection criterion
is almost equivalent to the optimal selection criterion for the
single relay selection.

Now let us calculate the BER expression of the S-RS-NC scheme using
the \textit{Min-Max} selection criterion. Let $\hat{b}_{r,1}(k)$ and
$\hat{b}_{r,2}(k)$ denote the estimation of  $b_{r}(k)$ at user 1's
and user 2's receivers, respectively. It can be easily verified that
any error in $\hat{b}_{r,j}(k), j=1,2$, will result in a bit error
at user $j's$ receiver. For example, if $\hat{b}_{r,1}(k)$ is in
error, then the estimation of  $b_2(k)$ at user 1's receiver, given
by $\hat{b}_2(k)=\hat{b}_{r,1}(k)\oplus b_1(k)$, will also be
erroneous, because user 1 knows $b_1(k)$ perfectly. This means that
the error probability of $b_2(k)$  at user 1's receiver is exactly
the same as the error probability of  $\hat{b}_{r,1}(k)$. If we let
$P_{r_i,u_j}$ be the BER in the link from relay $i$ to user $j$,
then the average BER of two users in this link, denoted by
$P_{r_i}$, is equal to $P_{r_i}=(P_{r_i,u_1}+P_{r_i,u_2})/2$. As
discussed before, it can also be approximated by using the BER of
the worst user as follows
\begin{equation} \label{eq8}
P_{r_i}=\frac{1}{2}(P_{r_i,u_1}+P_{r_i,u_2})=\frac{1}{2}max\{P_{r_i,u_1},P_{r_i,u_2}\}
\end{equation}

Let $\gamma_{ij}=\bar{p}_{r,i}/\sigma_n^2$ represent the average SNR
in the link from relay $i$ to user $j$. We also assume that all
$\gamma_{ij}$ for $i=1,\ldots,N$, $j=1,2,$ are the same and equal to
$\gamma_{rd}=\bar{p}_r/\sigma_n^2$, where $\bar{p}_r$ is the total
relay transmission power.

Let $P_{r_i,u_j}(\gamma_{rd}|h_{r_i,u_j}(k))$ represent the
instantaneous BER, in the link from relay $i$ to user $j$. It is
given by
\begin{eqnarray} \label{eq9}
P_{r_i,u_j}\left(\gamma_{rd}|h_{r_i,u_j}(k)\right)=Q\left(\sqrt{2\gamma_{i}^{u_j}(k)}\right)
\end{eqnarray}
where $\gamma_{i}^{u_j}(k)=\gamma_{rd}|h_{r_i,u_j}(k)|^2$ is the
instantaneous received SNR in the link from relay $i$ to user $j$.

If we let $z=\gamma_{i}^{u_j}(k)$, the PDF of $z$ is given by
\begin{eqnarray} \label{eq10}
f_Z(z)=\gamma_{rd}^{-1}exp\left(-z\gamma_{rd}^{-1}\right).
\end{eqnarray}

Let $P_{r_i,max}(\gamma_{rd}|h_{r_i,min}(k))=$ $\max
\biggl\{P_{r_i,u_1}(\gamma_{rd}|h_{r_i,u_1}(k))$
$,P_{r_i,u_2}(\gamma_{rd}|h_{r_i,u_2}(k))\biggl\}$ denote the error
probability of the worst user for the signals transmitted from relay
$j$, where
$|h_{r_i,min}(k)|=\min\left\{|h_{r_i,u_1}(k)|,|h_{r_i,u_2}(k)|\right\}$.
Then based on the \textit{Min-Max} selection criterion, among all
relays, a single relay, denoted by $S$, is selected such that the
instantaneous BER of the worst user is minimal. That is,
\begin{align} \label{eq11}
&S=\arg \min_i\{P_{r_i,max}(\gamma_{rd}|h_{r_i,min}(k))\} \\
&=\arg
\min_i\{max\{P_{r_i,u_1}({\gamma_{rd}|h_{r_i,u_1}(k)}),P_{r_i,u_2}({\gamma_{rd}|h_{r_i,u_2}(k))}\}
\nonumber \\ & \quad \quad \quad \quad \quad \quad \quad \quad \quad
\quad \quad \quad \quad \quad \quad \quad \quad \quad \quad
,i=1,\ldots,N\} \nonumber
\end{align}
and other unselected relays will stay in the idle states.

Since $Q(x)$ is a monotonic decreasing function of $x$, the
\textit{Min-Max} criterion in (\ref{eq11}) can be further written as
\begin{align} \label{eq12}
S &= \arg \max_i\{\gamma_{i}^{min}(k)\}  \\
&= \arg
\max_i\{min\{\gamma_{i}^{u_1}(k),\gamma_{i}^{u_2}(k)\},i=1,\ldots,N\}
\nonumber \\ &= \arg
\max_i\{min\{|h_{r_i,u_1}(k)|^2,|h_{r_i,u_2}(k)|^2\},i=1,\ldots,N\}
\nonumber
\end{align}
where
$\gamma_{i}^{min}(k)=min\{\gamma_{i}^{u_1}(k),\gamma_{i}^{u_2}(k)\}$
is the minimum of the two instantaneous SNRs of the links from relay
$i$ to the two users. That is
\begin{eqnarray} \label{eq13}
\gamma_{i}^{min}(k)&=&\min\{\gamma_{i}^{u_1}(k),\gamma_{i}^{u_2}(k)\}
\nonumber
\\ &=& \gamma_{rd}\min\{|h_{r_i,u_1}(k)|^2,|h_{r_i,u_2}(k)|^2\}.
\end{eqnarray}

Let $z_{min}=\gamma_{i}^{min}(k)$. From Eqs. (\ref{eq10}) and
(\ref{eq13}), the PDF of $z_{min}$ is given by\cite{14}
\begin{eqnarray} \label{eq14}
f_{Z_{min}}(z)=2\gamma_{rd}^{-1}exp(-2\gamma_{rd}^{-1}z).
\end{eqnarray}
Then the exact and asymptotic BER of the S-RS-NC scheme is given in
the following theorem.

{\bf{Theorem 1}} The average sum BER of the S-RS-NC scheme using the
\textit{Min-Max} selection criterion is given by
\begin{eqnarray} \label{eq15}
P^{S-RS-NC}(\gamma_{rd})\approx{\frac{1}{4}\sum_{p=0}^N{\binom{N}{p}(-1)^p\frac{1}{\sqrt{1+2p\gamma_{rd}^{-1}}}}}.
\end{eqnarray}
It can be further approximated at high SNR as
\begin{eqnarray} \label{eq16}
P^{S-RS-NC}(\gamma_{rd})\approx{\frac{2^{N-2}\Gamma\left(N+\frac{1}{2}\right)}{\sqrt{\pi}}\gamma_{rd}^{-N}+o\left(\gamma_{rd}^{-N}\right)}
\end{eqnarray}

 \textit{Proof}: See Appendix A.

From the above equation, we can see that a diversity order of $N$
can always be achieved by the proposed S-RS-NC scheme in a two-way
relay network with $N$ relay nodes.

Next, let us discuss the selection decision procedure for S-RS-NC.
Similar to \cite{31}, to perform selection at the relay nodes, each
relay needs to listen to the request-to-send (RTS) and the
clear-to-send (CTS) packets from two source nodes, respectively.
Based on that, each relay node estimates its channel power gains
from two source nodes. Then, a backoff timer is set to be inversely
proportional to the relaying channel quality
$\gamma_i^{min}(k)=min\{\gamma_i^{u_1}(k),\gamma_i^{u_2}(k)\}$. The
best relay with the largest $\gamma_i^{min}(k)$ and the smallest
backoff timer can occupy the channel first. Thus S-RC-NC based on
\textit{Min-Max} criterion can be implemented in a decentralized
way.

\begin{figure}
\centering
\includegraphics[width=0.5\textwidth]{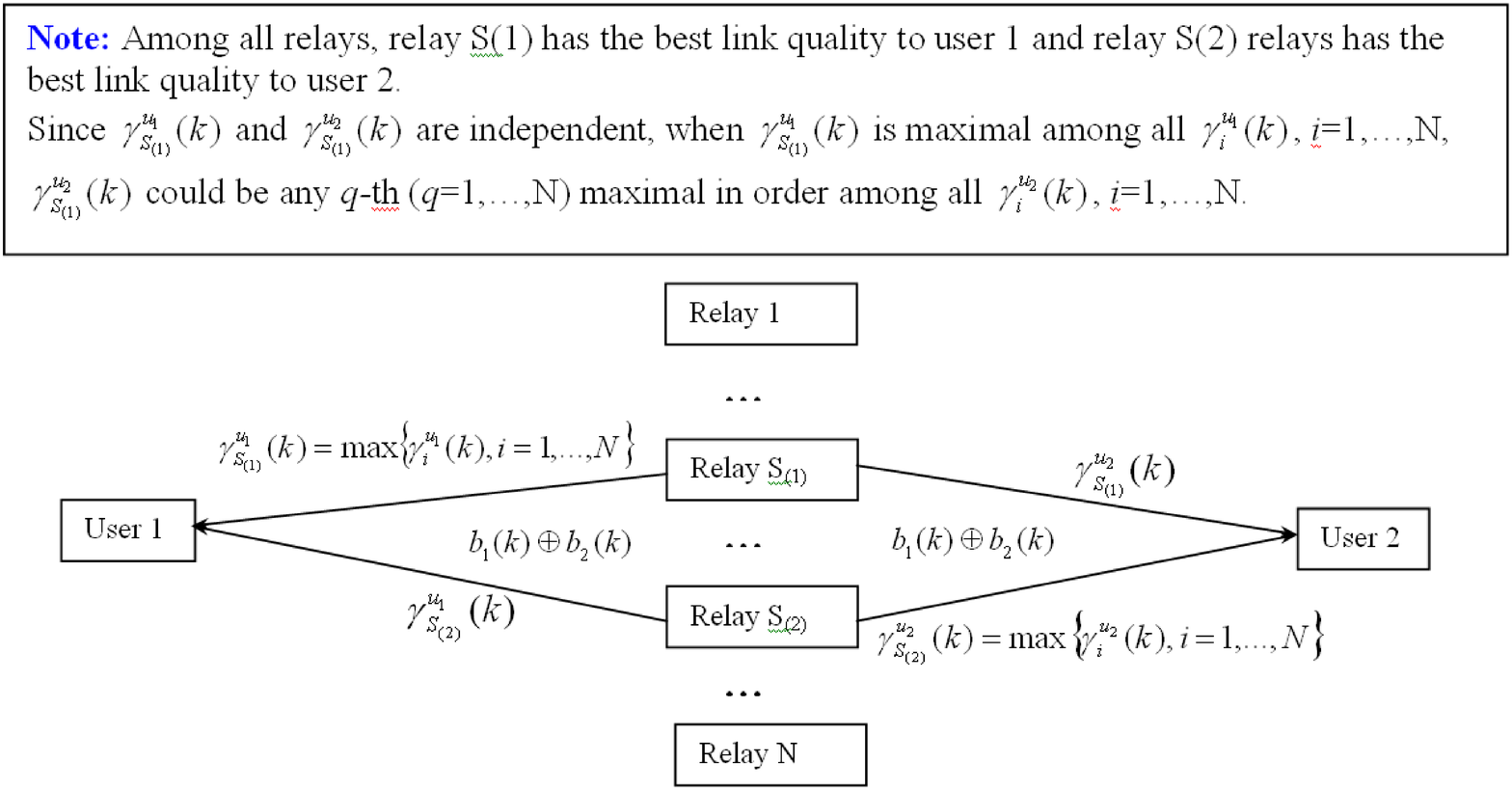}
\caption{Block diagram of D-RS-NC based on Double-Max selection
criterion} \label{fig:3}
\end{figure}

\begin{figure}
\centering
\includegraphics[width=0.5\textwidth]{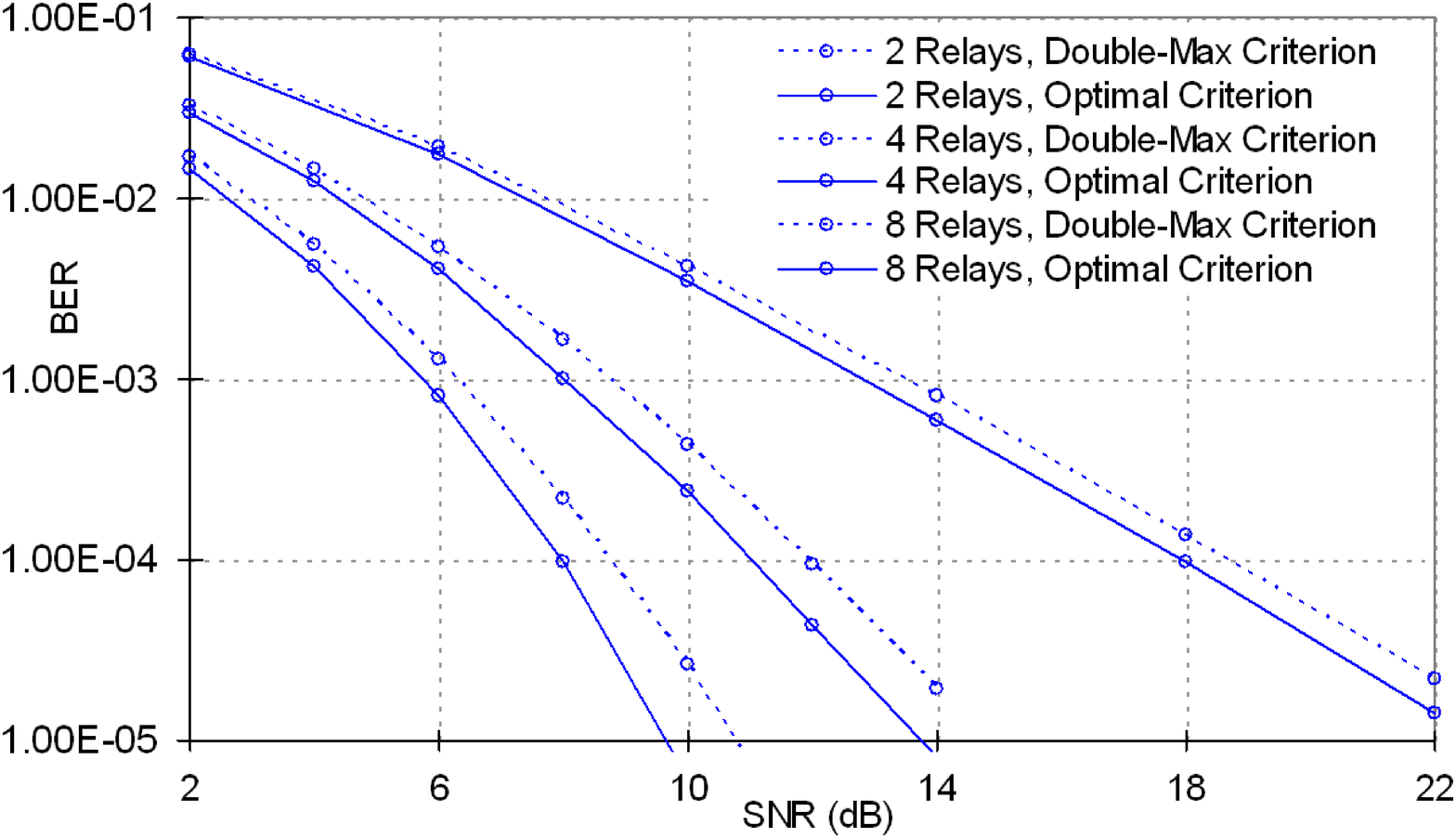}
\caption{BER performance comparison of D-RS-NC based on Double-Max
and optimal selection criterion} \label{fig:4}
\end{figure}

\subsection{Dual Relay Selection With Network Coding (D-RS-NC)}

In the S-RS-NC scheme, only a single relay is selected for
transmission. In this section, we consider another relay selection
algorithm, referred to as the dual-RS-NC (D-RS-NC). Again, our
selection criterion is to minimize the average sum BER of two users.
In the D-RS-NC, one or two out of $N$ relays are selected for
forwarding the network coded signals. Let  $S_{(1)}$ and $S_{(2)}$
denote the selected relays, which could be the same relay or
different relays. Furthermore, to simplify the relay selection
process and analysis, we consider a simple dual relay selection
algorithm, referred to as the \textit{Double-Max} selection
criterion, where we select one best relay for each user. Fig. 3
illustrates the relay selection process using the $Double$-$Max$
criterion. Specifically, we select two relays  $S_{(1)}$ and
$S_{(2)}$ such that relay  $S_{(j)}$ out of all relays has the best
link quality to the user $j$. That is, among all relays, the link
from relay $S_{(j)}$  to user $j$ has the maximum received SNR at
user $j, j=1, 2$,
\begin{eqnarray} \label{eq17}
S_{(j)}=\arg \max_i\{|h_{r_i,u_j}(k)|^2,i=1,\ldots,N\},j=1,2
\end{eqnarray}
That is,  $|h_{r_{S_{(j)}},u_j}(k)|$ is the maximum one among all
$|h_{r_i,u_j}(k)|$ for $i=1,\ldots,N$,
\begin{align} \label{eq18}
|h_{r_{S_{(j)}},u_j}(k)|^2=\max\{|h_{r_i,u_j}(k)|^2,i=1,\ldots,N\}
,j=1,2. \nonumber \\
\end{align}

Fig. 4 compares the D-RS-NC scheme based on the \textit{Double-Max}
criterion, with that based on the optimal criterion, which selects
optimal relays such that the instantaneous sum BER of two users is
minimized. The simulation setup is the same as in Fig. 2. For the
optimal criterion, for simplicity we also use a Chernoff bound to
approximate the $Q$-function when determining the optimal relay(s).
It can be seen that there is about 1dB gap between the
\textit{Double-Max} \textit{criterion} and the optimal selection
criterion at the BER of $10^{-4}$. However, as will be discussed
shortly, the \textit{Double-Max} scheme can be implemented in a very
simple decentralized way, but the optimal scheme needs to be
implemented in a centralized way. It requires one central node which
has channel state information (CSI) of all links to determine the
optimal relays minimizing the sum BER and this has to be done by an
exhaustive search among all possible pairs of relays. It requires
extensive overhead and high implementation complexity. By contrary,
the \textit{Double-Max} scheme is much simpler compared to the
optimal scheme and is only about 1dB away from the optimal scheme.
Also analysis of the optimal selection criterion involves the order
statistics of complex $Q$-functions which makes it difficult to
obtain closed-form expressions. Therefore, in this paper, we only
consider the \textit{Double-Max} criterion and derive both exact and
asymptotic BER expressions. This will give more insights into the
system performance.

Since the two selected relays transmit at the same time, they will
interfere with each other. To avoid interference, we consider using
an Alamouti space time block code (STBC) technique \cite{15} at the
two selected relays  $S_{(1)}$ and $S_{(2)}$ to facilitate
orthogonal transmission. Let $b_r(k)=b_1(k) \oplus b_2(k)$ and
$s_r(k)$ represent the modulated symbol of $b_r(k)$. Let
$s_r^{s_{(j)}}(k)$ represent the signals transmitted from relay
$S_{(j)}$ at time $k$. Then for STBC, the signal matrix transmitted
from $S_{(1)}$ and $S_{(2)}$ for the $k$-th and $k+1$-th symbols,
can be written as
\begin{eqnarray} \label{eq19}
\left[\begin{array}{cc}
s_r^{s_{(1)}}(k) & s_r^{s_{(1)}}(k+1)  \\
s_r^{s_{(2)}}(k) & s_r^{s_{(2)}}(k+1)
\end{array} \right] \qquad \qquad~~~~~~~ \nonumber \\ =\left[\begin{array}{cc}
\sqrt{\bar{p}_{r,1}}s_r(k)  & -\sqrt{\bar{p}_{r,2}}s_r^*(k+1)   \\
\sqrt{\bar{p}_{r,2}}s_r(k+1) & \sqrt{\bar{p}_{r,1}}s_r^*(k)
\end{array} \right]
\end{eqnarray}
where  $\bar{p}_{r,j}$, $j=1, 2$, represents the transmission power
at relay  $S_{(j)}$, $\bar{p}_{r,1}+\bar{p}_{r,2}=\bar{p}_{r}$ and
$\bar{p}_{r}$ is the total transmission power at the relays. In this
paper, we consider equal power allocation between two selected
relays. That is, $\bar{p}_{r,1}=\bar{p}_{r,2}=\bar{p}_{r}/2$.

We assume that the channel does not change during the two symbol
transmission duration. Following the same receiver process as in the
conventional Alamouti STBC system \cite{15}, the corresponding
received SNR at user $j$'s receiver, denoted by $\gamma^{u_j}(k)$,
can be calculated as
\begin{eqnarray} \label{eq20}
\gamma^{u_j}(k)=\frac{1}{2}\gamma_{rd}\left(|h_{r_{S_{(1)}},u_j}(k)|^2+|h_{r_{S_{(2)}},u_j}(k)|^2\right)
\end{eqnarray}
where $h_{r_{S_{(i)}},u_j}(k)$ is the fading coefficient from the
selected relay $S_{(i)}$ to user $j$. Then we have the following
theorem for the BER expression of D-RS-NC scheme.

{\bf{Theorem 2}} The average sum BER for the D-RS-NC using the
\textit{Double-Max} criterion is given by
\begin{eqnarray} \label{eq21}
P^{D-RS-NC}(\gamma_{rd}) &=&
\frac{1}{2}E\left(\sum_{j=1}^{2}Q\left(\sqrt{2\gamma^{u_j}(k)}\right)\right)
\nonumber \\ &=& \frac{\sqrt{\gamma_{rd}}}{2N} \sum_{p=0}^{N}
\binom{N}{p}(-1)^p \frac{1}{\sqrt{\gamma_{rd}+p}} \nonumber \\ &+&
(N-1)!(-1)^N
\sum_{q=1}^{N-1}\sum_{k=1}^{q}\sum_{p=q+1}^{N}\frac{\Psi_0}{\Sigma_0}
\nonumber
\end{eqnarray}
and it can be further approximated at high SNR as
\begin{eqnarray} \label{eq22}
P^{D-RS-NC}{(\gamma_{rd})}\approx \frac{
(2^N-1)\Gamma(N+\frac{1}{2})}{2N\sqrt{\pi}}\gamma_{rd}^{-N}+o(\gamma_{rd}^{-N})
\end{eqnarray}
where
\begin{small}
\begin{eqnarray}
\Sigma_0=\prod_{j=1,j\neq k}^{q}{(j-k)}\prod_{m=q+1, m\neq p}^{N}
(m-p)(N-P+1)(N-k+1), \nonumber
\end{eqnarray}
\begin{eqnarray} \label{eq22a}
\Psi_0=\biggl\{\begin{array}{cc}
\frac{1}{2}\left(1+\sqrt{\frac{c_1}{1+c_1}}\frac{c_1}{c_2-c_1}-\sqrt{\frac{c_2}{1+c_2}}\frac{c_2}{c_2-c_1}\right)
& c_1\neq c_2 \\
\frac{1}{2}\left(1-\sqrt{\frac{c_1}{1+c_1}}\left(1+\frac{1}{2(1+c_1)}\right)\right)
& c_1=c_2
\end{array},
\end{eqnarray}
\end{small}
\begin{eqnarray}
c_1=\frac{\gamma_{rd}}{N-k+1}, c_2=\frac{\gamma_{rd}}{2(N-p+1)}.
\nonumber
\end{eqnarray}

 \textit{Proof}: See Appendix B.

By comparing the asymptotic BER of D-RS-NC in Eq. (\ref{eq22}) with
the BER of S-RS-NC in Eq. (\ref{eq16}), we have
\begin{eqnarray} \label{eq23}
P^{D-RS-NC}(\gamma_{rd})=G_{D/S}P^{S-RS-NC}(\gamma_{rd})
\end{eqnarray}
where $G_{D/S}=\frac{2-2^{-(N-1)}}{N}<1$ for $N>1$ represents the
BER reduction of D-RS-NC relative to S-RS-NC.

Next, let us discuss the relay selection process for D-RS-NC. In the
D-RS-NC, each user only needs to select a best relay for itself, so
the decentralized selection method in the S-RS-NC scheme can also be
applied here, but the selection needs to be done in two steps. In
each step, one relay is selected for one user. Similar to S-RS-NC,
to select a best relay to user 1, each relay sets a backoff timer
which is inversely proportional to the relaying channel quality
$\gamma_{i}^{u_1}(k)$. The best relay with the largest
$\gamma_{i}^{u_1}(k)$  and the smallest backoff timer can occupy the
channel first and knows that it is the optimal relay to user 1.
Similarly, to select a best relay to user 2, each relay sets a
backoff timer which is inversely proportional to
$\gamma_{i}^{u_2}(k)$ and the optimal relay for user 2 can then be
selected. Thus D-RC-NC can also be implemented in a decentralized
way.

It can be observed from Eq. (\ref{eq23}) that unlike the
conventional relay networks, where a single relay selection is
optimum, in the two-way relay networks, the performance of
dual-relay selection with NC is always superior to the single relay
selection with NC.

\subsection{Network Coding without Relay Selection (NC -No-RS)}

As a comparison, let us now consider a conventional network coding
scheme with no relay selection (NC-No-RS). We assume that all relays
transmit on orthogonal channels, so that the receiver can separate
them without any interference from each other. This can be done by
orthogonal space time block coding or other orthogonal transmission
techniques. We should note that the NC-No-RS achieves the same
spectral efficiency as the NC with RS schemes. The STBC orthogonal
transmission in NC-No-RS is only used to facilitate the receiver
processing and it does not increase the frequency bandwidth and
reduce the spectral efficiency, compared to the NC with RS schemes.

For a fair comparison, we assume that the total relay transmission
power in the NC-No-RS scheme is the same as that in the RS-NC
schemes. In this paper, we assume that the power is equally
distributed among all relays.

After receiving $N$ independent signals from $N$ relays, each user's
receiver will then combine all the received signals. The overall
received SNR at the $j$-th user's receiver at time $k$, denoted by
$\gamma_{sum,j}(k)$, $j=1, 2$, can be calculated as
\begin{eqnarray} \label{eq24}
\gamma_{sum,j}(k)=\frac{\gamma_{rd}}{N}\sum_{i=1}^{N}{|h_{r_i,u_j}(k)|^2}
\end{eqnarray}
where $\gamma_{rd}=\bar{p}_r/N_0$.

{\bf{Theorem 3}} The average sum BER of NC-No-RS is given by
\begin{align} \label{eq25}
P^{NC-No-RS}(\gamma_{rd})=\frac{1}{2}-
\frac{1}{2\sqrt{\pi}}\sum_{p=0}^{N-1}
\frac{(N\gamma_{rd}^{-1})^p\Gamma(p+1/2)}{p!(1+N\gamma_{rd}^{-1})^{(p+1/2)}}.
\nonumber \\
\end{align}
Its asymptotic BER expression at high SNR can be approximated as
\begin{eqnarray} \label{eq26}
P^{NC-No-RS}(\gamma_{rd})\approx
\frac{N^N\Gamma(N+1/2)}{2\sqrt{\pi}\Gamma(N+1)}\gamma_{rd}^{-N}+o(\gamma_{rd}^{-N})
\end{eqnarray}

By comparing the asymptotic BER of NC-No-RS in Eq. (\ref{eq26}) with
the BER expression of S-RS-NC shown in Eq. (\ref{eq16}), we have
\begin{eqnarray} \label{eq27}
P^{S-RS-NC}(\gamma_{rd})\approx G_{S-RS}P^{NC-No-RS}(\gamma_{rd})
\end{eqnarray}
where $G_{S-RS}=\frac{N!2^{N-1}}{N^N}$ represents the BER reduction
of the S-RS-NC compared to the NC-No-RS. The gain comes from the
relay selection. It can be verified that $G_{S-RS}\leq1$ for $N>0$,
and it decreases with increasing $N$. This means that a relay
selection scheme is always superior to non-selection schemes with
all relay participation. The performance gain using relay selection
increases as the number of relay increases.

\subsection{Conventional Relay Selection With No Network Coding (RS-No-NC)}

As a comparison, now let us also consider a conventional relay
selection (RS) scheme without using network coding in two-way relay
networks. To avoid interference, we also assume that all the relays
transmit on orthogonal channels, so that they do not interfere with
each other. In the RS-No-NC scheme, we select one best relay for
each destination user and each selected relay will be used to
forward other user's information to its destination user.
Specifically, user 1 and user 2 first broadcast their own
information to all the relays, separately. One relay, which has the
best link quality to user 2 is selected for forwarding user 1's data
to user 2, while the another relay which has the best channel to
user 1 is selected for forwarding user 2's data to user 1. From the
above description, we can see that the RS-No-NC is somewhat similar
to the D-RS-NC scheme based on \textit{Double-Max} criterion.
Therefore, relay selection is the same as the D-RS-NC scheme. The
only difference is that there is no network coding in RS-No-NC
scheme and each selected relay only carries one user's information.

For a fair comparison, we also assume that the total transmission
power of two selected relays is equal to $\bar{p}_r$.  Following the
similar analysis as for the S-RS-NC scheme, the average BER of
RS-No-NC, denoted by $P^{RS-No-NC}(\gamma_{rd})$, can be calculated
as
\begin{eqnarray} \label{eq28}
P^{RS-No-NC}(\gamma_{rd})= \qquad \qquad \qquad \qquad ~\nonumber
\\
\frac{1}{2}E\left(Q\left(\sqrt{\gamma_{N,1}^{max}(k)}\right)+Q\left(\sqrt{\gamma_{N,2}^{max}(k)}\right)\right)
~\nonumber
\\ =
\frac{1}{2\sqrt{2\pi}}\int_0^{\infty}
\frac{e^{-x/2}}{\sqrt{x}}\left(1-e^{-\gamma_{rd}^{-1}x}\right)^Ndx
~~~ \nonumber \\ = \frac{1}{2\sqrt{2}}\sum_{p=0}^{N}
\binom{N}{p}(-1)^p\left(\frac{1}{2}+p\gamma_{rd}^{-1}\right)^{-1/2}
\end{eqnarray}
where
$\gamma_{N,j}^{max}(k)=\max\{\gamma_{i}^{u_j}(k),i=1,\ldots,N\}$ for
$j=1,2$ and $\gamma_{i}^{u_j}(k)=\gamma_{rd}|h_{r_i,u_j}(k)|^2$.

Again using the first order expansion of the PDF of
$\gamma_{N,j}^{max}(k)$ and Eq. (\ref{eq7}), its asymptotic BER at
high SNR is given by
\begin{eqnarray} \label{eq29}
P^{RS-No-NC}(\gamma_{rd})\approx
\frac{2^{N-1}\Gamma(N+1/2)}{\sqrt{\pi}}\gamma_{rd}^{-N}+o(\gamma_{rd}^{-N})
\end{eqnarray}

By comparing Eq. (\ref{eq29}) with Eq. (\ref{eq16}), (\ref{eq16})
can be rewritten as
\begin{eqnarray} \label{eq30}
P^{S-RS-NC}(\gamma_{rd})=\frac{1}{2}P^{RS-No-NC}(\gamma_{rd})
\end{eqnarray}

From the above equation, we can see that the error rate of the
S-RS-NC is always half of the conventional RS with no NC. This gain
is due to network coding.

Table 1 summarizes the BER reduction of various relay selection
schemes relative to the NC with no relay selections (NC-No-RS) and
they are illustrated in Fig. 5. They are calculated based on the
asymptotic BER at high SNRs, given in Eqs. (\ref{eq16}),
(\ref{eq22}), (\ref{eq26}) and (\ref{eq30}). It can be seen from the
table and figure that the D-RS-NC scheme has the maximum BER
reduction among all relay selection schemes, and thus performs best.
The S-RS-NC has the second best performance. The conventional RS
with no NC is worse than these two RS schemes, but is superior to
the NC with no relay selection as the number of relays is greater
than six.

\begin{table}
\renewcommand{\arraystretch}{1}
\caption{Gain of various relay selection algorithms relative to the
NC-No-RS} \label{tab:1} \centering
\begin{tabular}{|c|c|}
\hline Relay Selection Scheme & BER reduction relative to the
NC-No-RS \\ \hline S-RS-NC & $\frac{N!2^{N-1}}{N^N}$ \\ \hline
D-RS-NC & $\frac{(N-1)!(2^N-1)}{N^N}$ \\ \hline Conventional
RS-No-NC & $\frac{N!2^N}{N^N}$ \\ \hline
\end{tabular}
\end{table}

\begin{figure}
\centering
\includegraphics[width=0.5\textwidth]{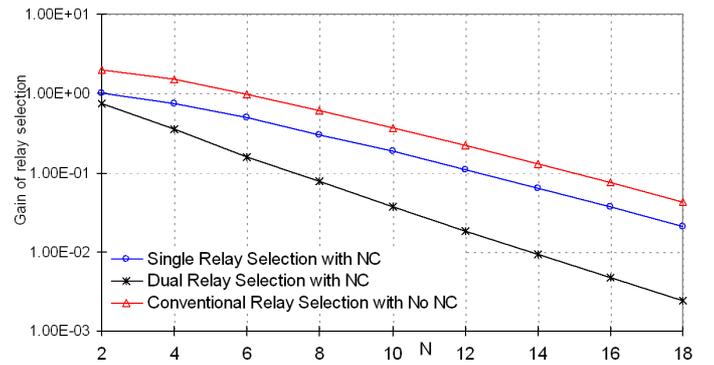}
\caption{BER reduction of various relay selection schemes relative
to the NC with no relay selection} \label{fig:5}
\end{figure}

\begin{figure}
\centering
\includegraphics[width=0.5\textwidth]{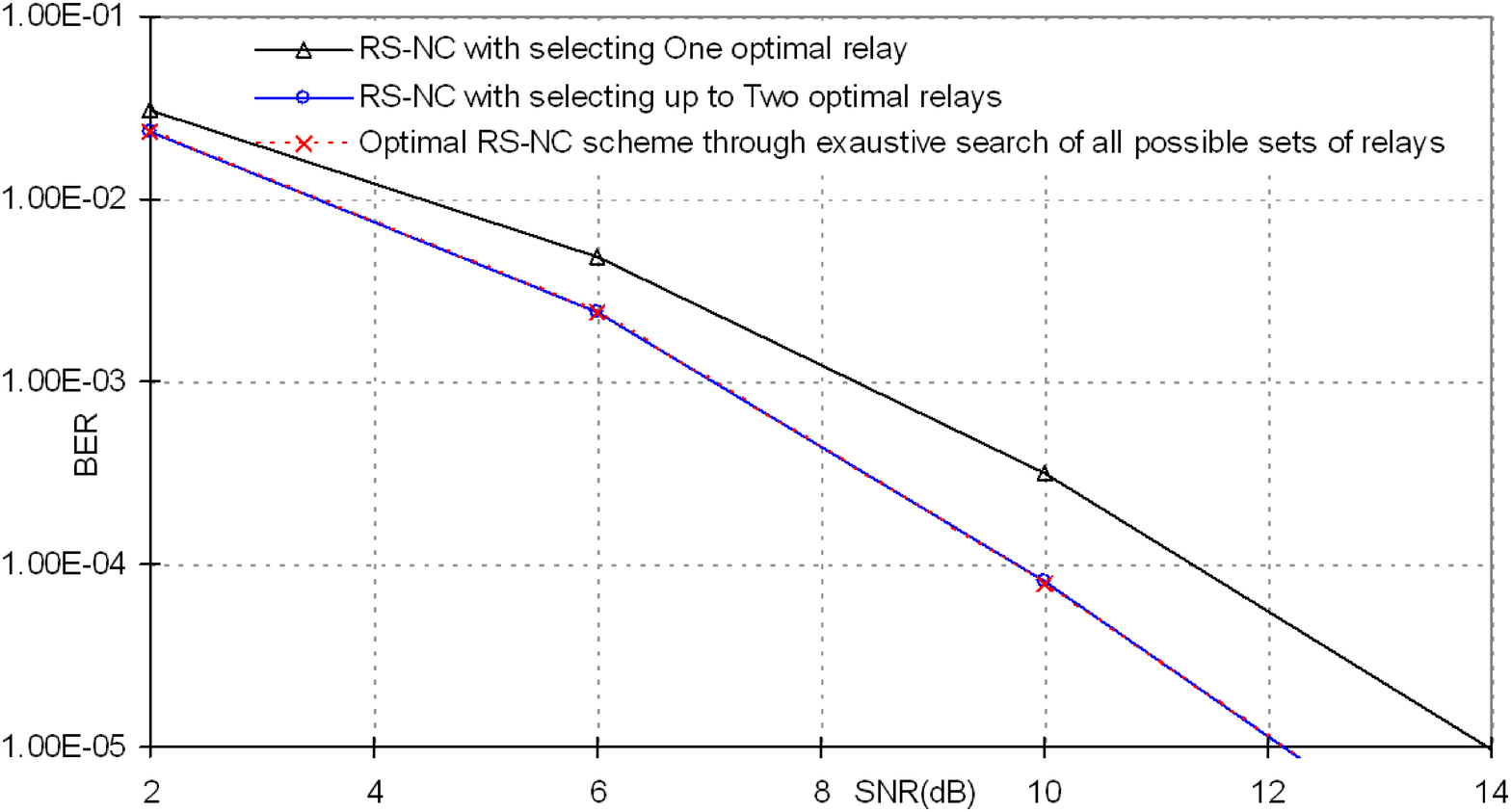}
\caption{BER performance comparison with selecting various numbers
of relays for 5 relays} \label{fig:6}
\end{figure}

\section{Simulation Results}

In this section, we provide simulation results. All simulations are
conducted using a BPSK modulation. We assume that the average SNR in
the link from relay $i$ to user $j$ for $i=1,\ldots,N$ and $j=1, 2$,
are all equal.  We consider the scenario where the link from the
source to relay is much more reliable than the link from the relay
to destination and as such the decoding errors at the relay can be
ignored.

In Fig. 6, we first compare the BER performance of RS-NC when
selecting various numbers of relays. The optimal RS-NC scheme is to
exhaustively search the optimal relay set among all possible sets of
relays, where the optimal relay set may contain one, two, three,
$\cdots$, up to $N$ relays. Obviously, such search has very high
complexity, especially when $N$ is very large. From the figure, we
can see that the D-RS-NC, which selects up to two optimal relays
performs almost the same as the optimal RS-NC scheme but has much
low complexity compared to the optimal relay set selection. This
indicates that the D-RS-NC scheme is a near optimal RS-NC scheme in
a two-way relay network. It also brings significant gains compared
to the S-RS-NC scheme.

Figs. 7-10 compares the BER performance of the single relay
selection with NC (S-RS-NC) based on \textit{Min-Max} criterion,
dual-relay selection with NC (D-RS-NC) based on \textit{Double-Max}
criterion, conventional relay selection with no NC (RS-No-NC) and
the NC with no relay selection (RS-No-NC) schemes for various
numbers of relays. It can be seen that among all these relay
selection schemes, D-RS-NC performs best. This is different from the
conventional one way relay networks, where the single relay
selection is optimal [11-13]. For the network with two relays, the
dual relay selection with NC can provide 0.5dB gain over the S-RS-NC
and NC-No-RS, and about 2dBs over the conventional RS-No-NC. The
gains are due to the relay selection and they may increase as the
number of relays increases. For example, as the number of relays is
increased to 16, the gain of the D-RS-NC over the S-RS-NC, RS-No-NC
and NC-No-RS are increased to 0.5dB, 1dB and 2.5dB, respectively.
This is consistent with the analytical results obtained in Section
III. However, the gain resulting from relay selection in two-way
networks is not as large as relay selection in one way relay
networks. This is due to the two-way communication models and use of
network coding.

\begin{figure}
\centering
\includegraphics[width=0.5\textwidth]{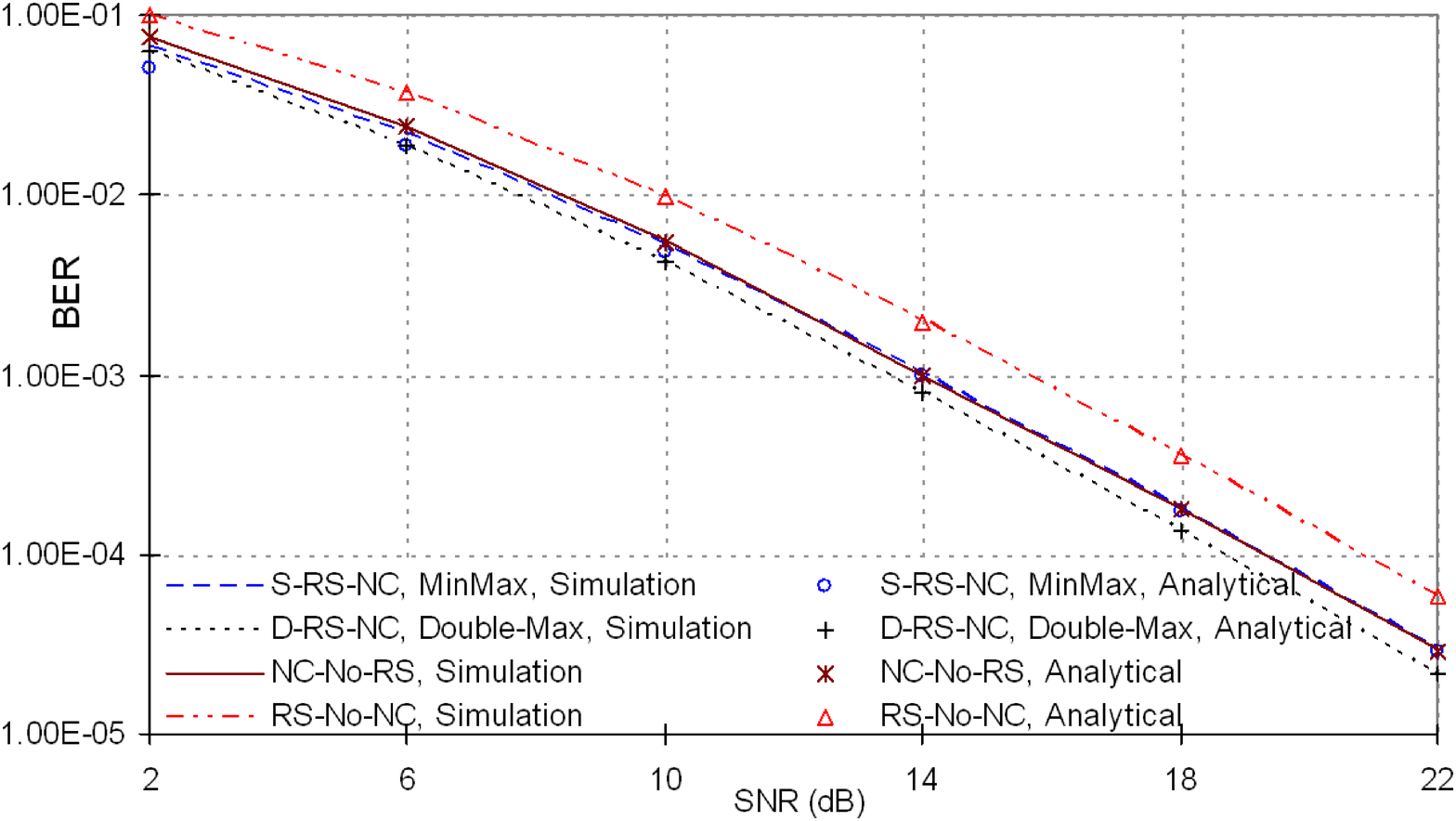}
\caption{BER performance comparison for 2 relays} \label{fig:7}
\end{figure}

\begin{figure}
\centering
\includegraphics[width=0.5\textwidth]{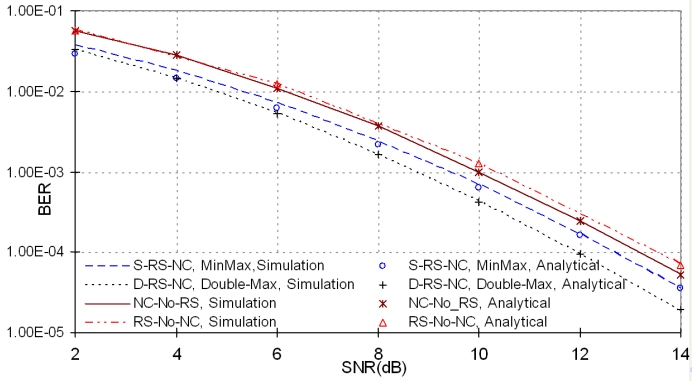}
\caption{BER performance comparison for 4 relays} \label{fig:8}
\end{figure}

Additionally, we can observe that the single relay selection with NC
does not provide any selection gain compared to the NC-No-RS when
the number of relays is two. Only when the number of relay is great
than two, it provides some gains. This is also different from the
relay selection in the conventional one way relay networks, where
the relay selection can provide a gain as long as there are more
than one relays.

Furthermore, we can also note that the combined relay selection with
network coding can provide reasonable gains compared to the
conventional relay selection with no network coding under the same
complexity. The gains are contributed from the network coding.

\begin{figure}
\centering
\includegraphics[width=0.5\textwidth]{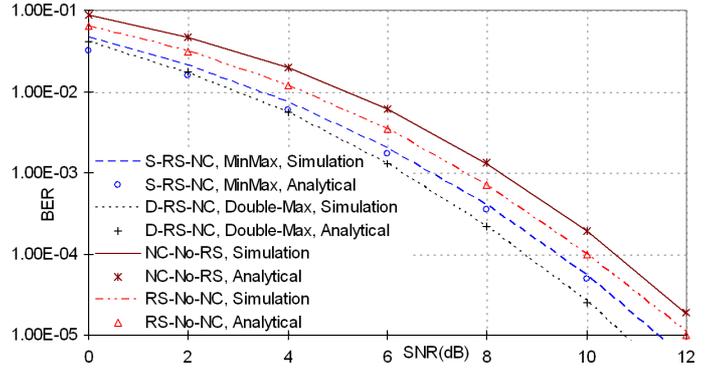}
\caption{BER performance comparison for 8 relays} \label{fig:9}
\end{figure}

\begin{figure}
\centering
\includegraphics[width=0.5\textwidth]{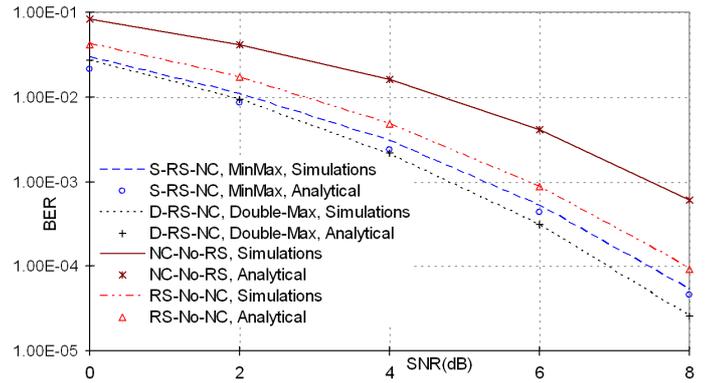}
\caption{BER performance comparison for 16 relays} \label{fig:10}
\end{figure}

Figs. 7-10 also compare the simulation results with the analytical
results using the exact BER expressions derived in Section III. It
can be seen that for the D-RS-NC, RS-No-NC, NC-No-RS schemes the
analytical BER exactly match the simulation results in all SNR
ranges. This validates that the BER expressions derived in Section
III are accurate. The only BER expression, which has slight
deviation from the simulations, is the S-RS-NC scheme. At low SNRs,
its analytical BER is slightly better than the simulation results.
This is because the analytical BER only uses the worst user's BER to
approximate the average sum BER of two users. However, as the SNR
increases, the analytical BER also matches the simulation results.
That is, it is accurate at high SNRs.

From the simulation results, we can see that unlike the conventional
relay networks, where a single relay selection is optimum, in
two-way relay channels, D-RS-NC outperforms other considered relay
selection schemes. However, the D-RS-NC requires strict time
synchronization between two selected relays. In contrast, S-RS-NC
does not have this requirement. Therefore, there are some
implementation complexity and performance tradeoff between these
schemes in practical applications.


%

\section{Conclusions}

In this paper, we combine the network coding and relay selection to
improve the transmission efficiency and system performance in
two-way relay channels. We proposed several selection schemes that
optimize the overall performance of two users, including single
relay selection with NC (S-RS-NC) and dual relay selection with NC
(D-RS-NC). To simplify the relay selection process, we propose
several simplified selection criteria. In particular, for S-RS-NC,
we proposed a \textit{Min-Max} selection criterion, so that the BER
of the worst user out of two users is optimized, while for D-RS-NC,
we developed a simple \textit{Double-Max} criterion, for which we
select one best relay for each user. Unlike the conventional relay
selection strategy in one way relay networks, where the optimal
selection strategy is to select a single best relay, in two-way
relay networks, the dual relay selection outperforms single relay
selection and is a near-optimal scheme. It was also shown that a
properly combined network coding and relay selection can improve the
system performance and efficiency. It provides additional selection
gains compared to pure NC schemes with no relay selection, and an
extra network coding gain relative to the conventional relay
selection with no NC, but the gain due to the relay selection in
two-way relay channels is not as larger as for the one way relay
channel. Furthermore, in terms of practical implementation, single
relay selection is easier to be implemented compared to the D-RS-NC
scheme as the later requires synchronization of relay transmission.
The performance of D-RS-NC using \textit{Double-Max} criterion is
about 1dB away from the optimal criterion. How to design a simple
selection criterion which can approach the optimal criterion would
be a very interesting problem.

\section{Appendix}

\subsection{Proof of Theorem 1}

Let $\gamma_S^{max-min}(k)=\max\{\gamma_{i}^{min}(k),i=1,\ldots,N\}$
=$\max \{ \min \{\gamma_{i}^{u_1}(k),\gamma_{i}^{u_2}(k)
\},i=1,\ldots,N \}$ represent the minimum value among the
instantaneous SNRs from the selected relay $S$ to users 1 and 2.
Based on Eqs. (\ref{eq12}) and (\ref{eq14}), the PDF of
$\gamma_S^{max-min}(k)$ can be calculated \cite{14}
\begin{eqnarray} \label{eq31}
f_{max}(x)&=&Nf_{Z_{min}}(x)(F_{Z_{min}}(x))^{N-1} \\
&=&2N\gamma_{rd}^{-1}exp(-2\gamma_{rd}^{-1}x)(1-exp(-2\gamma_{rd}^{-1}x))^{N-1}
\nonumber
\end{eqnarray} \label{eq32}
where $F_{Z_{min}}(x)$ is the CDF of $\gamma_{i}^{min}(k)$, given by
\begin{eqnarray}
F_{Z_{min}}(x)=1-exp(-2\gamma_{rd}^{-1}x)). \nonumber
\end{eqnarray}

The instantaneous overall BER from the selected relay $S$ to the two
users, denoted by $P_{r_S}(\gamma_{rd}|\gamma_S^{max-min})$, can be
approximated by
\begin{eqnarray} \label{eq33}
&& P_{r_S}(\gamma_{rd}|\gamma_S^{max-min}) \nonumber
\\ &=& \frac{1}{2}\{P_{r_S,u_1}(\gamma_{rd}|h_{r_S,u_1}(k))+P_{r_S,u_2}(\gamma_{rd}|h_{r_S,u_2}(k))\}
\nonumber
\\ &\approx&
\frac{1}{2}max\{P_{r_S,u_1}(\gamma_{rd}|h_{r_S,u_1}(k)),P_{r_S,u_2}(\gamma_{rd}|h_{r_S,u_2}(k))\} \nonumber \\
&=& \frac{1}{2}Q(\sqrt{2\gamma_S^{max-min}}). \nonumber
\end{eqnarray}

The average sum BER, denoted by $P^{S-RS-NC}(\gamma_{rd})$, can be
derived by averaging the above equation with respect to
$\gamma_S^{max-min}$ and is given by
\begin{eqnarray} \label{eq34}
&&
P^{S-RS-NC}(\gamma_{rd})\approx\frac{1}{2}E\left(Q\left(\sqrt{2\gamma_S^{max-min}}\right)\right)
\nonumber \\ &=& \frac{1}{4\sqrt{\pi}}\int_0^{\infty}
\frac{e^{-x}}{\sqrt{x}}F_{max}(x)dx \nonumber \\
&=& \frac{1}{4}\sum_{p=0}^N
\binom{N}{p}(-1)^p\frac{1}{\sqrt{1+2p\gamma_{rd}^{-1}}}
\end{eqnarray}
where we have made use of Eq. (\ref{eq5}) with $b=2$ and
$F_{max}(x)=\left(1-e^{-2x\gamma_{rd}^{-1}}\right)^N$ is the CDF of
$\gamma_S^{max-min}$.

This yields Eq. (\ref{eq15}).

At high SNR, its first order expansion is given by
\begin{eqnarray} \label{eq34}
F_{max}(x)=\left(1-e^{-2x\gamma_{rd}^{-1}}\right)^N=\left(2x\gamma_{rd}^{-1}\right)^N+o(x^{N+\epsilon}),0<\epsilon<1
\nonumber
\end{eqnarray}

By simply algebraic manipulation of the above equation to obtain the
PDF, and using Eq. (\ref{eq7}), we can obtain its asymptotic BER as
shown in Eq. (\ref{eq16}).

\subsection{Proof of Theorem 2}

We assume that all fading coefficients are independent. Let
$\gamma_{i}^{u_j}(k)=\gamma_{rd}|h_{r_i,u_j}(k)|^2$. For each user
$j$, $j=1,2$, we rearrange $\gamma_{i}^{u_j}(k)$, $i=1,\ldots,N$, in
an ascending order, denoted by $\gamma_{(i)}^{u_j}(k)$, such that
$\gamma_{(1)}^{u_j}(k)\leq$ $\gamma_{(2)}^{u_j}(k)\leq$ $\cdots$
$\leq$ $\gamma_{(N)}^{u_j}(k)$ $=\gamma_{S_{(j)}}^{u_j}(k)$, $j=1,
2$. That is, we arrange $\gamma_{(i)}^{u_j}(k)$ as the $(N-i+1)$-th
maximal in order. As we can see from Fig. 3, since
$\gamma_{S_{(j)}}^{u_1}(k)$ and $\gamma_{S_{(j)}}^{u_2}(k)$ are
independent, when $\gamma_{S_{(1)}}^{u_1}(k)$ is the maximal among
all $\gamma_{i}^{u_1}(k)$, $\gamma_{S_{(1)}}^{u_2}(k)$ could be any
$q$-th ($q=1,\ldots,N$) maximal in order among all
$\gamma_{i}^{u_2}(k)$. That is, for any $q$, $q=1,\ldots,N$, the
probability that $\gamma_{S_{(1)}}^{u_2}(k)$ is the $q$-th maximal
in order is the same and equal to $1/N$. Similarly, when
$\gamma_{S_{(2)}}^{u_2}(k)$ is the maximum one among all
$\gamma_{i}^{u_2}(k)$, $\gamma_{S_{(2)}}^{u_1}(k)$ could be of any
$q$-th maximal in order among all $\gamma_{i}^{u_1}(k)$.

The average BER for D-RS-NC, denoted by
$P^{D-RS-NC}{(\gamma_{rd})}$, can be calculated as
\begin{eqnarray} \label{eq35}
&&
P^{D-RS-NC}{(\gamma_{rd})}=\frac{1}{2}E\left(\sum_{j=1}^{2}Q\left(\sqrt{2\gamma^{u_j}(k)}\right)\right)
 \nonumber \\
&=&
E\left(Q\left(\sqrt{\gamma_{S_{(1)}}^{u_1}(k)+\gamma_{S_{(2)}}^{u_1}(k)}\right)\right)
\nonumber \\
&=&
\frac{1}{N}\sum_{S_{(1)}=S_{(2)}}E\left(Q\left(\sqrt{\gamma_{S_{(1)}}^{u_1}(k)+\gamma_{S_{(2)}}^{u_1}(k)}\right)\right)
 \qquad  \nonumber \\ &+& \frac{1}{N}\sum_{S_{(1)}\neq
S_{(2)}}E\left(Q\left(\sqrt{\gamma_{S_{(1)}}^{u_1}(k)+\gamma_{S_{(2)}}^{u_1}(k)}\right)\right) \nonumber \\
&=&
\frac{1}{N}\sum_{S_{(1)}=S_{(2)}}E\left(Q\left(\sqrt{2\gamma_{(N)}^{u_1}(k)}\right)\right)
 \nonumber \\ &+& \frac{1}{N}\sum_{S_{(1)}\neq
S_{(2)}}E\left(Q\left(\sqrt{\gamma_{(N)}^{u_1}(k)+\gamma_{(q)}^{u_1}(k)}\right)\right)
\nonumber \\ &=& L_1+L_2
\end{eqnarray}

The first term $L_1$ can be calculated by following the similar
procedure as for the S-RS-NC and is given by
\begin{eqnarray} \label{eq36}
L_1 &=&
\frac{1}{N}\sum_{S_{(1)}=S_{(2)}}E\left[Q\left(\sqrt{2\gamma_{(N)}^{u_1}(k)}\right)\right]
\nonumber \\ &=& \frac{\sqrt{\gamma_{rd}}}{2N}\sum_{p=0}^N
\binom{N}{p}(-1)^p\frac{1}{\sqrt{p+\gamma_{rd}}}
\end{eqnarray}

Now let us calculate the second term $L_2$. We first find the moment
generating function (MGF) of
$Z=X_1+X_2=\gamma_{(N)}^{u_1}+\gamma_{(q)}^{u_1}$. Let us define
$w_l=\gamma_{(l)}^{u_1}-\gamma_{(l-1)}^{u_1}$ for $l=2,\ldots,N$ and
$w_1=\gamma_{(1)}^{u_1}$. It can be shown that $w_l$ for
$l=1,\ldots,N$ are independent, and distributed according to
\cite{23}
\begin{eqnarray} \label{eq37}
f_{w_l}(w_l)=\frac{N-l-1}{\gamma_{rd}}exp\left(-\frac{N-l-1}{\gamma_{rd}}w_l\right)
\end{eqnarray}

Then we have
\begin{eqnarray} \label{eq38}
Z &=& \gamma_{(N)}^{u_1}+\gamma_{(q)}^{u_1}=\sum_{k=1}^N w_k +
\sum_{k=1}^q w_k \nonumber \\ &=& 2\sum_{k=1}^q w_k + \sum_{k=q+1}^N
w_k
\end{eqnarray}

The moment generate function (MGF) of $Z$ is thus given by
\begin{align}\label{eq39}
M_Z(s) &= \int_0^\infty \ldots \int_0^\infty \left(\prod_{k=1}^{N}f_{w_k}(w_k)\right) e^{sZ} {\rm d} w_1 \ldots {\rm d} w_N \notag \\
&=   \left(\prod_{k=1}^q \int_0^\infty  e^{2s w_k} f_{w_k}(w_k) {\rm d} w_k \right) \nonumber \\ & \left(\prod_{k=q+1}^N \int_0^\infty  e^{s w_k} f_{w_k}(w_k) {\rm d} w_k\right) \notag \\
&=   \frac{N! (-1)^N }{2^q \gamma_{rd}^N} \prod_{k=1}^q \frac{1}{s
-\frac{ N-k+1}{2 \gamma_{rd}}}  \prod_{k=q+1}^N \frac{1}{s-\frac{
N-k+1 }{\gamma_{rd}} } \; .
\end{align}

By using the partial fraction expansion, Eq. (\ref{eq39}) can be
further expressed in Eq. (\ref{eq40}).
\begin{figure*}
\begin{small}
\begin{align} \label{eq40}
M_Z(s) &= \frac{N! (-1)^N }{2^q \gamma_{rd}^N}  \sum_{k=1}^q \frac{1}{ \left(s - \frac{N-k+1}{2 \gamma_{rd}}\right) \prod_{j=1, j \neq k}^q\left(\frac{N-k+1}{2 \gamma_{rd}}-\frac{N-j+1}{2 \gamma_{rd}} \right)} \notag  \sum_{p=q+1}^N \frac{1}{ \left(s - \frac{N-p+1}{\gamma_{rd}}\right) \prod_{j=q+1, j \neq p}^{N}\left(\frac{N-p+1}{\gamma_{rd}}-\frac{N-j+1}{\gamma_{rd}} \right)} \notag \\
&= \frac{N! (-1)^N }{2 \gamma_{rd}^2}  \sum_{k=1}^q \sum_{p=q+1}^N  \frac{1}{  \prod_{j=1, j \neq k}^q\left(j-k\right) \prod_{m=q+1, m \neq p}^{N}\left(m-p\right)} \frac{1}{ \left(s - \frac{N-k+1}{2 \gamma_{rd}}\right) \left(s - \frac{N-p+1}{\gamma_{rd}}\right) } \notag \\
\end{align}
\end{small}
\end{figure*}

Based on Eq. (\ref{eq40}), $L_2$ can then be calculated in Eq.
(\ref{eq41}),
\begin{figure*}
\begin{small}
\begin{align} \label{eq41}
L_2&= \frac{1}{N}\sum_{S_{(1)} \neq S_{(2)}} {\rm E} \left[ Q\left(\sqrt{\gamma_{(N)}^{u_1}+\gamma_{(q)}^{u_1}} \right) \right] \notag \\
&= \sum_{q=1}^{N-1} \frac{(N-1)! (-1)^N }{2 \gamma_{rd}^2}  \sum_{k=1}^q \sum_{p=q+1}^N  \frac{1 }{  \prod_{j=1, j \neq k}^q\left(j-k\right) \prod_{m=q+1, m \neq p}^{N}\left(m-p\right)}\notag  \frac{1}{\pi} \int_0^{\frac{\pi}{2}} \frac{1}{ \left(\frac{1}{2 \sin^2 (\theta)} +  \frac{N-k+1}{2 \gamma_{rd}}\right) \left(\frac{1}{2 \sin^2 (\theta)} + \frac{N-p+1}{\gamma_{rd}}\right) } {\rm d} \theta \notag \\
&= \sum_{q=1}^{N-1}  (N-1)! (-1)^N \sum_{k=1}^q \sum_{p=q+1}^N
\frac{1 }{  \prod_{j=1, j \neq k}^q\left(j-k\right) \prod_{m=q+1, m
\neq p}^{N}\left(m-p\right)(N-p+1)(N-k+1)} \Psi_0
\end{align}
\end{small}
\end{figure*}
where $\Psi_0= \frac{1}{\pi}\int_0^{\frac{\pi}{2}} \frac{\sin^4
(\theta)}{ \left(c_1 +  \sin^2 (\theta)\right) \left(c_2 + \sin^2
(\theta)\right) } {\rm d} \theta $ and it can be further calculated
in (\ref{eq22a}).

By substituting (\ref{eq41}) and (\ref{eq36}) into (\ref{eq35}), we
can obtain the desired expression in Eq. (\ref{eq21}).

To derive the high SNR approximation, we first evaluate the first
order expansion of the PDF of $Z=X_1+X_2$ by applying a Taylor
series expansion around the origin. This can be written as
\begin{eqnarray} \label{eq43}
f_Z(z)=\frac{f_Z^{(p)}(0)z^p}{p!}+o(z^{p+1})
\end{eqnarray}
where $f_Z^{(p)}(0)$ is the $p$-th derivative of $f_Z(z)$ with
respect to $z$, evaluated at zero, and $p$ is the minimum integer
value such that $f_Z^{(p)}(0)\neq 0$. Using the derivative property
of the Laplace transform and the Initial Value Theorem, we can
restate the equivalent problem as finding the minimum value of $p$
such that
\begin{eqnarray} \label{eq44}
f_Z^{(p)}(0)=lim_{s\rightarrow \infty} s^{p+1}M_Z(-s) \neq 0
\end{eqnarray}

By substituting Eq. (\ref{eq39}) into Eq. (\ref{eq44}), we can show
that $p=N-1$, which gives
\begin{eqnarray} \label{eq45}
f_Z(z)=N2^{-q}z^{N-1}+o(z^N)
\end{eqnarray}

Substituting Eq. (\ref{eq45}) into Eq. (\ref{eq7}), the asymptotic
BER at high SNR can be written as
\begin{eqnarray}
P^{D-RS-NC}(\gamma_{rd})=\frac{\Gamma(N+\frac{1}{2})}{2N\sqrt{\pi}}\gamma_{rd}^{-N}
\nonumber \\ +\sum_{q=1}^{N-1}\frac{2^{N-1}N\Gamma(N+\frac{1}{2})2^{-q}}{\sqrt{\pi}N^2}\gamma_{rd}^{-N}+o(\gamma_{rd}^{-N})  \nonumber \\
=\frac{
(2^N-1)\Gamma(N+\frac{1}{2})}{2N\sqrt{\pi}}\gamma_{rd}^{-N}+o(\gamma_{rd}^{-N})
\end{eqnarray}
This proves Theorem 2.

\subsection{Proof of Theorem 3}
Let $F_{\gamma_{sum,j}}(x)$ represent the CDF of $\gamma_{sum,j}$,
then it is given by
\begin{eqnarray} \label{eq46}
F_{\gamma_{sum,j}}(x)=\frac{\gamma\left(N,Nx\gamma_{rd}^{-1}\right)}{\Gamma(N)}=1-e^{-Nx\gamma_{rd}^{-1}}\sum_{p=0}^{N-1}
\frac{(Nx\gamma_{rd}^{-1})^p}{p!} \nonumber
\end{eqnarray}

The average BER of NC-No-RS can then be calculated as
\begin{eqnarray} \label{eq47}
&& P^{NC-No-RS}(\gamma_{rd}) = \frac{1}{2}E\left(\sum _{j=1}^2
Q\left(\sqrt{2\gamma_{sum,j}(k)}\right)\right) \nonumber
\\ &=&\frac{1}{2}\sum_{j=1}^2 \int_0^{\infty} Q\left(\sqrt{2x_j}\right)
f_{\gamma_{sum,j}}(x_j)dx_j
\nonumber \\ &=& \frac{1}{2\sqrt{\pi}} \int_0^{\infty} \frac{e^{-x}}{\sqrt{x}}F_{\gamma_{sum,j}}(x)dx \nonumber \\
&=&
\frac{1}{2}-\frac{1}{2\sqrt{\pi}}\sum_{p=0}^{N-1}\frac{(N\gamma_{rd}^{-1})^p}{p!}\Gamma(p+\frac{1}{2})(1+N\gamma_{rd}^{-1})^{-(p+\frac{1}{2})}
\nonumber
\end{eqnarray}
This proves Theorem 3.

\bibliographystyle{IEEE}

\begin{thebibliography}{31}

\bibitem{1} A. Sendonaris, E. Erkip, and B. Aazhang, "User
cooperation diversity -Part I: system description," \textit{IEEE
Trans. Commun.}, vol. 51, pp. 1927-1938, Nov. 2003.

\bibitem{2}
R. Ahlswede, N. Cai, S Li, and R. W. Yeung, "Network information
flow," \textit{IEEE Trans. Inf. Theory}, vol. 46, no. 4, pp.
1204-1216, Jul.2000.


\bibitem{4}
J. N. Laneman, D. N. C. Tse and G. W. Wornell, "Cooperative
diversity in wireless networks: Efficient protocols and outage
behavior," \textit{IEEE Trans. Inform. Theory}, vol. 50, no. 12,
Dec. 2004, pp. 3062-2080.

\bibitem{5}

A. Madsen and J. Zhang, "Capacity bounds and power allocation for
wireless relay channels," \textit{IEEE Trans. Infom. Theory}, vol.
51, no. 6, pp. 2020-2040, Jun. 2005.

\bibitem{6}

C. Hausl, and J. Hagenauer, "Iterative network and channel decoding
for the two-way relay channel," \textit{Proc of ICC}, June 2006, pp.
1568 - 1573.

\bibitem{7}
P. Popovski, and H. Yomo, "Physical network coding in two-way
wireless relay channels," \textit{Proc of ICC}, June 2007, pp. 707 -
712.

\bibitem{8}
F. Xue, C.H Liu and S. Sandhu, "MAC-layer and PHY-layer network
coding for two-way relaying: achievable rate regions and
opportunistic scheduling," \textit{Proc of Allerton Conference
2007}, Sept 2007, pp. 396-402.

\bibitem{9}
I. Baik and S. Y. Chung, "Network coding for two-way relay channels
using lattices," \textit{Proc of ICC}, May 2008, pp. 3898 - 3902.


\bibitem{10}
R. Madan, et al., "Energy-efficient cooperative relaying over fading
channels with simple relay selection," \textit{IEEE Trans. Wireless
Commun.}, vol. 7, no. 8, pp.  3013 - 3025, Aug. 2008.

\bibitem{11}
Y. Zhao, R. Adve, and T. J. Lim, "Improving amplify-and-forward
relay networks: optimal power allocation versus selection,"
\textit{IEEE Trans. Wireless Commun.}, vol. 6, No. 8, pp. 3114 -
3123,  Aug. 2007.


\bibitem{12}
R. Madan, et al., "Efficient cooperative relaying over fading
channels with simple relay selection," \textit{IEEE Trans. Wireless
Commun.}, vol. 7, No. 8, pp. 3013 - 3025, Aug. 2008.

\bibitem{13}
Y. Li, B. Vucetic, Z. Chen and J. Yuan, "An improved relay selection
scheme with hybrid relaying protocols," \textit{Proc of GLOBECOM}.
Nov. 2007, pp. 3704 - 3708.


\bibitem{14}
H. A. David, \textit{Order Statistics}, \textit{Jonh Wiley \& Sons,
Inc}, 1970.


\bibitem{15}
S. M. Alamouti, "A simple transmit diversity technique for wireless
communications", \textit{IEEE JSAC}, vol.16, no.8, pp.1451-1458, Oct
1998.

\bibitem{16}
X. Bao and J. Li, "A unified channel-network coding treatment for
user cooperation in wireless Ad-Hoc networks," \textit{Proc of ISIT
2006}, July 2006, pp. 202 - 206.


\bibitem{17}
M. Yu, J. Li, and R. S. Blum, "User cooperation through network
coding," \textit{Proc of ICC 2007}, June 2007, pp. 4064 - 4069.


\bibitem{18}
P. Larsson, N. Johansson, K. E. Sunell, "Coded bi-directional
relaying," \textit{Proc of VTC 2006 Spring}, vol. 2, 2006, pp. 851 -
855.

\bibitem{19}
S. W. Kim, "Concatenated network coding for large-scale multi-hop
wireless networks," \textit{Proc of WCNC 2007}, Mar 2007, pp. 985 -
989.

\bibitem{20}
S. Zhang, Y. Zhu, S. Liew and K. Letaief, "Joint design of network
coding and channel decoding for wireless networks," \textit{Proc of
WCNC 2007}, Mar. 2007, pp. 779 - 784.


\bibitem{21}
C. Hausl and P. Dupraz, "Joint network-channel coding for the
multiple-access relay channel," \textit{Proc of SECON}. 2006, vol.
3, Sept. 2006, pp. 817 - 822.

\bibitem{22}
T. J. Oechtering, and H. Boche, "Bidirectional regenerative
half-duplex relaying using relay selection," \textit{IEEE Trans.
Wireless Commun.}, vol. 7, pp. 1879 - 1888, May 2008.


\bibitem{23}
M. K. Simon and M. Alouini, \textit{Digital communication over
fading channels}, 2nd edition, \textit{Jonh Wiley \& Sons}, Inc,
2005.

\bibitem{24}
Z. Wang and G.B. Giannakis, "A simple and general parameterization
quantifying performance in fading channels," \textit{IEEE Trans.
Commun.}, vol.51, no.8, pp.1389-1398, Aug. 2003.


\bibitem{25}

J. Yuan, et al., "Distributed space-time trellis codes for a
cooperative system," \textit{IEEE Trans Wireless Commun.}, vol. 8,
no. 10, pp. 4897-4905, Oct 2009.


\bibitem{26}

C. Wang, et al. "Superposition-coded concurrent decode-and-forward
relaying," \textit{Proc of ISIT 2008}, July 2008, pp. 2390 - 2394.

\bibitem{27}
S. X. Ng, Y. Li and L. Hanzo, "Distributed turbo trellis coded
modulation for cooperative communications," \textit{Proc of ICC
2009}, June 2009.

\bibitem{28}
F. A. Onat, Y. Fan, H. Yanikomeroglu1 and H. V. Poor, "Threshold
based relay selection in cooperative wireless networks,"
\textit{Proc of IEEE Globecom}, Nov. 2008.

\bibitem{29}
M. Eslamifar, C. Yuen, W. H. Chin, Y. L. Guan, "Max-Min Antenna
Selection for Bi-Directional Multi-Antenna Relaying",\textit{ Proc
of VTC-Spring 2010}, available at
http://www1.i2r.a-star.edu.sg/~cyuen/publications.html.


\bibitem{30}
M. Eslamifar, W. H. Chin, C. Yuen, and Y. L. Guan, "Performance
Analysis of Two-Way Multiple-Antenna Relaying with Network Coding",
\textit{Proc of VTC-Fall 2009}, 2009.

\bibitem{31}
A. Bletsas, A. Khisti, D. P. Reed, and A. Lippman, "A simple
cooperative diversity method based on network path selection,"
\textit{IEEE JSAC}, vol. 24, no. 3, pp. 659-672, Mar. 2006.


\end{thebibliography}

%




\end{document}